\documentstyle[12pt,epsf]{article}
\def\ba{\begin{eqnarray}\samepage}
\def\ea{\end{eqnarray}}
\input amssym.def
\input amssym.tex

\newcommand{\po}{Poincar\'{e} }

\def\nin{\in\hskip-0.35cm/\hskip0.1cm}
\newcommand{\mod}{\mbox{ mod\,}}
\newcommand{\siot}{$\mbox{sio}_2$\ }
\newcommand{\sioo}{$\mbox{sio}_1$\ }

\newcommand{\diff}{\partial}

\newcommand{\be}{\begin{equation}}
\newcommand{\ee}{\end{equation}}
\newcommand{\ben}{\begin{eqnarray}\displaystyle}
\newcommand{\een}{\end{eqnarray}}

\def\beq{\begin{equation}}
\def\eeq{\end{equation}}

\begin{document}

\title{Generalised  supersymmetry and  $p$-brane actions}
\author{Stephen Hewson\footnote{sfh10@damtp.cam.ac.uk}\\ 
Department of Applied Mathematics and Theoretical
Physics\\Silver Street, Cambridge, CB3 9EW\\England}
\maketitle

\begin{abstract}
We investigate the most general $N=1$  graded extension of the \po
algebra, and find the corresponding supersymmetry transformations and
the associated superspaces. We find that the supersymmetry for which
$\{Q,Q\}\sim P$ is not special, and in fact must be treated
democratically with a whole class of supersymmetries. We show that
there are  two distinct types of grading, and a new class of general
spinors is defined. The associated superspaces  are shown to be either
of the usual type,  or flat with no torsion. $p$-branes are discussed in
these general superspaces and twelve  dimensions emerges as
maximal. New types of brane are discovered which could explain many
features of the standard $p$-brane theories. 
\end{abstract}

\section{Introduction}
\setcounter{equation}{0}
It is becoming apparent that twelve dimensions may have an important role to
play in the formulation of theories of extended objects, and there
have been many suggestions as to the possible form of this higher
dimensional connection \cite{twelve_dimensions,ftheory}. Although
generalising 
theories of 
extended objects to higher dimensions is a simple matter from the
bosonic viewpoint, when one considers supersymmetry the
picture becomes rather  complex. Indeed, the relationship
between the ten and eleven  dimensional superstring and supermembrane theories
is well understood, at 
least classically, in part  due to the similarity of
supersymmetry in these two dimensions. As one moves from ten and
eleven dimensions to twelve
dimensions, however,  the supersymmetry algebras and the symmetry
properties of 
products of gamma matrices  change, and therefore the basic  extensions  used
to generalise Green-Schwarz strings to Green-Schwarz membranes are  no
longer   applicable: A more detailed study of the supersymmetry is
required, whence it is clear that in order to formulate
explicit twelve dimensional theories, the notion of supersymmetry
must be generalised. 
Using this generalised approach it was shown in 
\cite{hp:ftheory} that it was possible to define an
$N=1$ 
supersymmetric 
(2+2)-brane action in a spacetime of signature $(10,2)$. This brane,
however, is just one solution to one particular generalisation of
supersymmetry, and the question remains as to how many different new
$p$-branes and how many different new types of supersymmetry can be
defined, in {\it any} given dimension. Such a question shows that
there are 
gaps in the understanding of `supersymmetry', and the purpose of this
paper is fill in these  gaps by studying  the most general \po 
supersymmetries (a \po supersymmetry being a relationship between
 bosonic variables transforming as representations of the \po
group, and an additional set of fermionic degrees of freedom). The work
shall be in essentially two parts.  We shall first formulate
generalised supersymmetry theories \footnote{Very recently, some
extended aspects of 
supersymmetry have been discussed in the literature \cite{generalised_supersymmetry}}from a purely algebraic point of
view and find the associated generalised
superspaces. We will then formulate invariant objects in 
these superspaces and write down invariant actions. We will 
finally  consider a generalisation of $p$-brane
theory to include branes propagating in such general superspace backgrounds,
thus presenting new types of fundamental supersymmetric  objects.

\subsection{Background}
Historically, a theory  deemed to be
`$N=1$ supersymmetric' is defined on a supermanifold which has a group of
isometries locally described  by the following fermionic extension of the
bosonic \po algebra 
\ben\label{poincare}
[M_{\mu\nu},M_{\rho\sigma}]&=& 
M_{\nu\sigma}\eta_{\mu\rho} +M_{\mu\rho}\eta_{\nu\sigma}-M_{\nu\rho}\eta_{\sigma\mu}- M_{\sigma\mu}\eta_{\nu\rho} \,,\nonumber \\
\left[M_{\mu\nu},P_\rho\right]&=&P_\mu\eta_{\nu\rho}-P_\nu\eta_{\mu\rho}\,,\nonumber
\\
\left[M_{\mu\nu},Q^\alpha\right]&=&-\frac{1}{2}{(\Gamma_{\mu\nu})^\alpha}_\beta
Q^\beta\,,\nonumber \\ 
\left[P_\mu,P_\nu\right]&=&\left[P_\mu,Q\right]=0 \,,\nonumber\\
\{Q^\alpha,Q^\beta\}&=&(\Gamma_\mu)^{\alpha\beta}P^\mu\,,
\een
where $P_\mu$ and $M_{\mu\nu}$ are the usual translation and rotation
generators of the isometry group of flat $(S,T)$ space and $Q^\alpha$
is a spinorial 
generator. Of course, a supersymmetry is a way to include fermionic
terms to the usual bosonic notion of spacetime. Experimentally we know that
the bosonic part of space should locally be described by the \po
algebra; the inclusion of fermions to such a picture is, at present,  rather more
arbitrary. 

The supersymmetry algebra (\ref{poincare}) is the basis to the whole
subject of supersymmetric string theory \cite{gsw}, which developed
into the more general picture of $p$-branes. The $p$-branes can be thought of either as
fundamental branes with a Green Schwarz action, much like the 
superstring, or as  extended solitonic solutions to the corresponding
supergravity theory. These supergravity $p$-branes must couple to the
theory via  $p$-form central charges, $Z_{\mu_1\dots\mu_p}$,  which must
be added as 
extensions to the supersymmetry algebra (\ref{poincare}) as follows
\be
\{Q,Q\}=\Gamma^\mu
P_\mu+\Gamma^{\mu_1\dots\mu_p}Z_{\mu_1\dots\mu_p}\,.
\ee
Studies of such extensions of the \po algebra have lead to the
standard  picture of which $p$-branes are and are not possible; this,
of course, is all dependent on choosing (\ref{poincare}) as the
starting point. The question should be asked as to whether we find any
new branes if this basic starting point is questioned. 

As we shall explain in this paper, the usual  choice of supersymmetry
algebra (\ref{poincare}) is actually in no way singled out; we could
just as well choose our 
initial algebra to be 
\be
\{Q,Q\}=\Gamma^{\mu_1\dots\mu_p}P_{\mu_1\dots\mu_p}\,,
\ee
and extend with a central $Z^\mu$ if we wish to define a 1-brane. Such
a statement greatly enlarges the notion of supersymmetry, but as we
shall see it is possible to classify all \po supersymmetry theories
into just two classes: those for which $\{Q,Q\}\sim
\Gamma^{\mu\nu}M_{\mu\nu}+\dots$, and those for which it does not. 
In this paper we will first discuss the formulation of a generalised
\po supersymmetric theory, and we shall then consider the invariant
actions for $p$-branes propagating in such spaces. A new brane scan
will then be presented for all branes for which an action exists.

\subsection{Charge conjugation matrix}
Central to the notion of supersymmetry is the Clifford algebra which
may be  represented by the matrices
$\{{(\Gamma_\mu)^{\alpha}}_\beta\}$ as 
\be\label{clifford}
{\{\Gamma_\mu,\Gamma_\nu\}^\alpha}_\beta=2\eta_{\mu\nu}{(I)^\alpha}_\beta\,,
\ee
where $\eta=\mbox{diag}(-1,\dots,-1,1,\dots,1)$ is the flat space metric for which there are $T$
minus signs  and $S$ plus
signs, corresponding to timelike and spacelike directions respectively. 
The spinorial indices are raised and lowered with the charge
conjugation matrix $C$ and its inverse $C^{-1}$ respectively. The
properties of the charge conjugation vary according to the signature
of the spacetime under consideration. For this reason we present a
brief discussion of these matrices before we begin.

For a
Majorana representation, that is a purely real representation
corresponding to the Clifford algebra generated over the real numbers
\footnote{Given a real representation of the $(S,T)$ Clifford algebra,
we may always define a purely imaginary representation of the $(T,S)$
Clifford algebra. Throughout this paper we shall choose our gamma
matrices always to be real.}, $C$ has
the properties \cite{nieuwenhuizen,kt} that
\ben\label{gammaid}
{\tilde{\Gamma}}_\mu&=&(-1)^{T}\eta C\Gamma_\mu C^{-1}\nonumber \\
C^{\dag}C&=&1\nonumber \\
\tilde{C}&=&\epsilon\eta^T(-1)^{T(T+1)/2}C\,,
\een
where the tilde denotes matrix transpose. 
 The possible choices of the numbers
$\eta, \epsilon=\pm 1$ 
depend on the signature of the spacetime as follows
\be\label{eta}
\begin{array}{c|c|c}  
\epsilon &\eta&(S-T)\mbox{ mod}\,8\\ \hline
+1&+1&0,1,2\\ \hline
+1&-1&0,6,7\\ \hline
-1&+1&4,5,6\\ \hline
-1&-1&2,3,4\\ 
\end{array}
\ee
If $\tilde{\Gamma}_\mu=\pm C\Gamma_\mu
C^{-1}$ then we  define $C=C_\pm$. This definition must be 
 consistent with the conditions \cite{nieuwenhuizen}
\be\label{ctilde}
\tilde{C}_+=(-)^{T+[(T+1)/2]}C_+,\hskip1cm
\tilde{C}_-=(-)^{[(T+1)/2]}C_-\,, 
\ee
where the brackets denote `integer part'. 
We find that these constraints imply that $\epsilon=1$.   This gives us that
\be\label{tab}
\begin{array}{c|c|c} 
T\mod 4&(C,\pi(C)):\eta=1&(C,\pi(C)):\eta=-1\\ \hline
0&(C_+,1)&(C_-,1)\\ \hline
1&(C_-,-1)&(C_+,1)\\ \hline
2&(C_+,-1)&(C_-,-1)\\ \hline
3&(C_-,1)&(C_+,-1)\\ 
\end{array}
\ee
where $\pi$ is the parity operator for which $\pi(C)=(-)1$ for $C$
(anti)symmetric.

\section{Superspaces}
\setcounter{equation}{0}

The algebra (\ref{poincare}) has the defining property that it reduces to the
bosonic \po algebra when the spinor generator $Q_\alpha$ is set to
zero. There are, however, many consistent $N=1$ graded  extensions of
the \po algebra with such a  property. A general (Dirac) spinor in $D$
spacetime 
dimensions had $2^{[D/2]}$ complex components, where the square brackets
denote `integer part'. A Majorana spinor has $2^{[D/2]}$ real
components. The anticommutator $\{Q^\alpha,Q^\beta\}$ is thus a
symmetric $2^{\left[D/2\right]}\times 2^{\left[D/2\right]}$ matrix. In general, since
$\{\Gamma^{\mu_1\dots \mu_p}\equiv
\Gamma^{[\mu_1}\dots\Gamma^{\mu_p]}\}$, for $p=1\dots D$, form a basis
for the vector space
of 
real $2^{\left[D/2\right]}\times 2^{\left[D/2\right]}$ matrices,  the
$\{Q^\alpha,Q^\beta\}$  
anticommutator in expression (\ref{poincare}), may  most generally be
rewritten as  
\be\label{qq}
\{Q^\alpha,Q^\beta\}=\sum_{n\in k}
(\Gamma_{\mu_1\dots\mu_n})^{\alpha\beta} 
Z^{\mu_1\dots\mu_n}\,,
\ee
where $k$ is the set of all $n$  such that
$(\Gamma_{\mu_1\dots\mu_n})^{\alpha\beta}$ is symmetric in the spinor
indices. The $Z^{\mu_1\dots\mu_n}$ are bosonic generators which are
completely antisymmetric in the spacetime indices. 
If we consider the \po algebra as being the infinite radius limit of the
de Sitter algebra, which differs from the \po algebra by the term 
$\left[P_\mu,P_\nu\right]=mM_{\mu\nu}$
where $m^{-1}$ is the radius of the de Sitter space \cite{holten_proeyen}, 
then we find that if we choose to identify  $Z^{\mu\nu}\sim M^{\mu\nu}$
and $Z^{\mu}\sim P^{\mu}$ in the infinite radius limit, that 
\ben\label{zz}
[Z^{\mu_1\dots\mu_p},M^{\mu\nu}]&=&\eta^{[\mu_1|\mu} Z^{\nu|\mu_2\dots\mu_p]}-\eta^{[\mu_1|\nu}Z^{\mu|\mu_2\dots\mu_p]}\hskip1cm
\mbox{for } p=1\dots D\,\nonumber\\
\left[Z^{\mu_1\dots\mu_p},Z^{\nu_1\dots\nu_p}\right]&=&\left[Z^{\mu_1\dots\mu_p},Q^{\alpha}\right]=0\,,
\een
and we see that the $Z^{(p)}$ terms are added to the algebra in
exactly the same way as $Z^{\mu}=P^\mu$, unless $p=2$. In de Sitter
space there will in general be non-trivial commutation relations
between the $Z^{(p)}$ and the $Z^{(q)}$, for $p,q\neq 2$. In the \po
space limit, however, these additional commutation relations all
vanish. 
We thus see then  the $M^{\mu\nu}$ term is  singled out from the
$\{Z^{(i)}\}$: all others appear on an equal footing regarding \po
supersymmetries.  This is an interesting point since it implies that  taking
$\{Q,Q\}\sim P$ as  in (\ref{poincare})  is {\it not} a natural
choice: all the other $Z^{(i)}$ appear with equal importance  as $P$
in the superalgebras. Hence, the whole subject of supersymmetric
$p$-branes is viewed from a biased position. 
It is for this reason that it is natural to
consider other types of supersymmetry theory. This is an  extension of
the $p$-brane democracy idea 
\cite{tow:pbranedem} to that of a `superspace democracy'.  By
exploring general forms of the 
anticommutator, we hope to find new properties  of supersymmetric
theories, and the relationships between them. This should then be
intrinsically linked to the existence of the different types of
$p$-branes.

 We shall consider the algebra with the graded extension
\be\label{newqq}
\{Q^\alpha,Q^\beta\}=\sum_{\tilde{n}\in\tilde{k}}
(\Gamma_{\mu_1\dots\mu_{\tilde{n}}})^{\alpha\beta} 
Z^{\mu_1\dots\mu_{\tilde{n}}}\,,
\ee
for any  $\tilde{k}\subseteq k$, $k$ being the full set of symmetric
matrices.  We shall call the \po extensions  with these
anticommutators $\mbox{sio}_{\tilde{k}}$. The $Z^{\mu_1\dots\mu_n}$
will commute with everything except $M^{\mu\nu}$, as in (\ref{zz}). 
It should be noted, of course, that one cannot generally expect these
supersymmetry algebras to be completely consistent from the the point
of view of the super-Jacobi identities. 
We shall assume that the algebras will be consistent on-shell, as is  the
case for the $\mbox{sio}_1$ supersymmetry algebra (\ref{poincare}). The aim
is to describe $p$-branes moving in any $D$-dimensional $\mbox{sio}_k$
invariant background.   In order to do this in a supersymmetric
fashion, we need to 
determine the superspaces corresponding to the algebras
$\mbox{sio}_{k}$, as was done in \cite{hp:ftheory} for the 12
dimensional $\mbox{sio}_2$ case.

We shall begin with a description of the usual case of $\mbox{sio}_1$.  This superspace is defined to be a coset
manifold ${\cal{G}}/{\cal{H}}$ where ${\cal{G}}$ is some supergroup
corresponding to the super-algebra $\mbox{sio}_1$, and ${\cal{H}}$
some 
subgroup of ${\cal{G}}$ \cite{dewitt}. Locally we may write $g\in{\cal{G}}$ as
\be\label{g}
g(X,\theta,\omega)=\exp(X.P+\theta.Q)\exp\left(\frac{1}{2}\omega.M\right)\,,
\ee
where $(X,\theta)$ are the superspace coordinates, and $\omega$ labels
each coset. The effect  of an infinitesimal group
action on the coordinates is found by multiplication on the left of $g$ by the infinitesimal group element
\be 
 g(\delta X,\delta\theta,0)=\exp(\delta
X.P+\delta\theta.Q)\,.
\ee
To reduce the product $\delta g.g$ to an element of the form of
(\ref{g}) we use the Baker-Campbell-Hausdorff formula, 
\be\label{bchformula}
\exp(\alpha A)\exp(B)=\exp\left(B+\alpha A +
\alpha\sum_{n=1}^\infty\frac{1}{(n+1)!}[\underbrace{[\dots[A,B],B],\dots,B}_{
\mbox{{n 
times}}}]+O(\alpha^2)\right)\,,
\ee
where $\alpha$ is an infinitesimal super-number. 
Application of this formula leads to the appearance of contractions of
the term $\{Q^\alpha,Q^\beta\}$ 
 in the exponential, which  produces the usual $\mbox{sio}_1$ supersymmetry
transformations
\be\label{suptrans1}
\delta\theta^\alpha=\epsilon^\alpha\,,\hskip1cm \delta
X^\mu=x^\mu-\frac{1}{2}\delta\theta_\alpha(\Gamma^\mu)^{\alpha\beta}\theta_\beta\,,
\ee
where $x^\mu$ and $\epsilon^\alpha$ are infinitesimal bosonic and
fermionic superspace
parameters respectively.

 The $\mbox{sio}_2$ supersymmetry transformations  may be  found
for the case that
$\{Q^\alpha,Q^\beta\}=\frac{1}{2}(\Gamma_{\mu\nu})^{\alpha\beta}M^{\mu\nu}$
in a 
similar way. However, since the anticommutator term in the algebra
generates a rotation, the left action of an infinitesimal group
element destroys the coset form of the group  unless the  following
restriction on the spinors is made:
\be\label{projection}
\delta\theta_\alpha(\Gamma^{\mu\nu})^{\alpha\beta}\theta_\beta.M_{\mu\nu}=0\,.
\ee
This identity will be found whenever the anticommutator of the
spinorial generators generates $M_{\mu\nu}$. 
In this situation, the corresponding supersymmetry transformations are
trivial
\be\label{suptrans2}
\delta X^\mu=x^\mu,\hskip1cm \delta\theta^\alpha=\epsilon^\alpha\,,
\ee
and the superspace is flat with no torsion. This is referred to as
`simple supersymmetry' \cite{freund_kaplansky}. 
 
The coset procedures just detailed are very general and may be
used to generate spaces which are labelled by other types of
parameters. These will be investigated in the next section, and will
all  be
shown to be of essentially  $\mbox{sio}_1$ or  $\mbox{sio}_2$ type.

\subsection{$\mbox{sio}_{\tilde{k}}$ supersymmetry}

We have presented  generalisations of the \po supersymmetry algebra for
which the translation generator $P^\mu$ is not
singled out. We can think of  these algebras as generating the
isometries of some background superspace, which will in general be
parametrised by the coordinates
$\{\theta^\alpha,X^\mu,X^{\mu\nu},\dots X^{\mu_1\dots\mu_D}\}$. We can
consider the generalised \po supersymmetric $p$-branes as objects propagating
in these new backgrounds. In order to 
do this we must apply the coset procedure to produce the generalised
superspaces. We shall perform the procedure for the  algebras
$\mbox{sio}_{\tilde{k}}$, which have the 
anticommutators (\ref{newqq}), to find that the general 
supersymmetry transformations are given by a combination of terms of
the type (\ref{suptrans1}) and (\ref{suptrans2}). Note that the
general coset theory which we shall employ works in exactly the same way
as for usual non-supersymmetric groups and manifolds, at least in the
case where the body of the supergroup, the part which remains when we
set the spinorial  terms $Q^\alpha=0$,   is itself a Lie algebra
\cite{dewitt}, which is the case of interest to us.

We can now write the superspace as a coset
${\cal{G}}/{\cal{H}}$. Until now we have naturally considered any
super-extension of the bosonic \po algebra. In order to keep the coset 
procedure natural, the question must now be asked as to which
subgroup ${\cal{H}}$ should we choose to quotient
by. Since every bosonic generator has a non-trivial commutation of
qualitatively the same form with $Z^{\mu\nu}$,     it is natural to
quotient out by this term. We can then write down the supergroup
elements parametrised by  $\{\omega^{\mu\nu};X^{p}:p\in
P\subset\{1,3,4,\dots,D\}\}$ for each $P$ as  as 
\be\label{generalg}
g(X,\theta,\omega)\equiv\exp \left(\sum_{p\in
P}X^{\mu_1\dots\mu_p} Z_{\mu_1\dots\mu_p}+\theta^\alpha
Q_\alpha\right)\exp\left(\frac{1}{2}\omega^{\mu\nu}M_{\mu\nu}\right)\,,
\ee
Notice that we allow any choice of the set $P$, even if it includes
values of $p\nin \tilde{k}$.  This is an important point: Even though  the
spinorial terms do not generate a particular $p$-form, there is no
reason not to include it as a parameter of the underlying
supermanifold; we simply add the purely bosonic commutator
$\left[Z^{(p)},M\right]\sim Z^{(p)}$ to the algebra.  The
coordinates of the superspaces invariant under the action of
$\mbox{sio}_{\tilde{k}}$ will  be given by
$\{\{X^{\mu_1\dots\mu_p}\},\theta^\alpha\}$, where $p\in
P$; the
cosets will be labelled by the parameters $\omega_{\mu\nu}$. 

We now evaluate the effects of the left action of an infinitesimal
group element which is constant in each coset, in that
$\delta\omega_{\mu\nu}=0$, on (\ref{generalg}). We require that the
supermanifold have isometries generated by the appropriate superalgebra,
hence that the supermanifold should be invariant under such a left
shift. 
After simplification using the BCH formula, we find that 
\ben
 g(\delta X,\delta\theta,0)
g(X,\theta,\omega)
&=&\exp\left( 
\delta X.Z+X.Z +(\delta\theta+\theta)^\alpha
Q_\alpha+C+D\right)\exp\left(\frac{\omega^{\mu\nu}M_{\mu\nu}}{2}\right)\nonumber\\
C&=&\frac{1}{2}[\delta X.Z+\delta \theta^\alpha
Q_\alpha,X.Z+\theta^\beta
Q_\beta]\nonumber\\
X.Z&\equiv & \sum_{p\in P}X^{\mu_1\dots\mu_p}Z_{\mu_1\dots\mu_p}\,,
\een
where the $[\,\,,\,\,]$ denotes the  super-commutator. The $D$
term is formed by repeated commutation of the exponent of $g(\delta X,\delta\theta,0)$
with $C$. In order to evaluate these terms we  refer to the
super-commutation relations (\ref{newqq}) and 
(\ref{zz}), from which it is clear that
\be\label{c}
C=\delta\theta^\alpha\{Q_\alpha,Q_\beta\}\theta^\beta=\frac{1}{2}\delta\theta^\alpha\sum_{p\in
{\tilde{k}}}
(\Gamma_{\mu_1\dots\mu_p})_{\alpha\beta} 
Z^{\mu_1\dots\mu_p}\theta^\beta\,.
\ee
If the anticommutator of the supersymmetries does not generate the
rotations $M_{\mu\nu}$  then $D$ is zero, since $[C,\delta
X.Z+\delta\theta^\alpha Q_\alpha]=0$. We can now read off the
supersymmetry transformations as being the shift in the coordinates
induced by the infinitesimal transformation. 
In this case, we therefore find
that the action of the supergroup yields the following supersymmetry
transformations
\be\label{susytrans}
\delta\theta^\alpha=\epsilon^\alpha\,,\hskip1cm\delta
X^{\mu_1\dots\mu_p}=x^{\mu_1\dots\mu_p}+{\sigma_p}\epsilon_\alpha(\Gamma_{\mu_1\dots\mu_p})^{\alpha\beta}\theta_\beta\,\,\,\,\,\,\,\forall
p\in P\,,
\ee
where $\sigma_p=1$ if ${p}\in \tilde{k}$ and $\sigma_p=0$
otherwise. Hence if $Z^{(p)}$ is generated by $\{Q,Q\}$ then
$\sigma_p=1$, and the supersymmetry transformation is of an
exactly analogous form to the usual  supersymmetry for the $p=1$
case. If  $Z^{(p)}$ is not generated by $\{Q,Q\}$ then
$\sigma_p=0$, and the corresponding supersymmetry transformation is
trivial, as we would expect.

In the situation where the anticommutation of the supersymmetries does
generate $M_{\mu\nu}$ we find that $D\neq 0$ in general, since
the commutation of $C$ with $Q$ produces $M_{\mu\nu}$, which does not
trivially commute with any of the generators. This destroys the coset
construction since we are unable to factor out the $M$-dependence to
give an expression of the form
\be
 g(\delta X,\delta\theta,0)
g(X,\theta,\omega)
=\exp\left( 
\tilde{X}.Z+\tilde{\theta}.Q\right)\exp\left(\frac{\tilde{\omega}^{\mu\nu}M_{\mu\nu}}{2}\right)\,,
\ee
for some $\tilde{X},\tilde{\theta},\tilde{\omega}$. 
The only way to solve this problem is to impose the constraint
(\ref{projection}) that $C=0$:  
\be\label{proj}
\delta\theta_\alpha(\Gamma^{\mu\nu})^{\alpha\beta}\theta_\beta M_{\mu\nu}=0\,\,\,\,\,\,\forall
\mu,\nu\,.
\ee
Of course, since $\delta\theta$ is an arbitrary infinitesimal spinor
we require that the equation must be satisfied for {\it any general pair of
spinors} $\psi,\phi$. If we do not wish to place any restriction on
the $M_{\mu\nu}$ then we must have  
\be\label{projection11}
\psi_\alpha(\Gamma^{\mu\nu})^{\alpha\beta}\phi_\beta=0\,\,\,\,\,\,\forall
\mu,\nu\,.
\ee
 This is the defining relation of the $\mbox{sio}_2$ superspace spinors, or
indeed any superspace for which $2\in \tilde{k}$: The expression
(\ref{projection11}) must hold for 
the coset construction to be well defined.  
If the identity is satisfied, then the supersymmetry transformations
(\ref{susytrans}) are unchanged, and 
(\ref{projection11}) merely serves to restrict the number of spin
variables. 

We are now in a position to calculate the $\mbox{sio}_{\tilde{k}}$
invariant forms from which we may build invariant actions.   From the 
transformations (\ref{susytrans}) we can 
write down forms which are invariant under the action of the
supergroup $\mbox{SIO}_{\tilde{k}}$, and therefore are supersymmetric
\be\label{oneforms}
\Pi^\alpha=d\theta^\alpha\,,\hskip1cm
\Pi^{\mu_1\dots\mu_p}=dX^{\mu_1\dots \mu_p}+{\sigma_p}\theta_\alpha(\Gamma_{\mu_1\dots\mu_{p}})^{\alpha\beta}d\theta_\beta\,,
\ee
where $d$ is the exterior derivative. This shows that there are in
fact precisely two general forms of supersymmetry transformation associated
with the \po group: those for which $\sigma_p=0$ and those for
which $\sigma_p=1$. The $\sigma_p=0$ cases, those for which
$\{Q,Q\}\nsim Z^{(p)}$,  are essentially the trivial
cases, corresponding to a flat supersymmetry with no torsion. 
If, however, the superspace is of $\mbox{sio}_2$ type, for which
$\{Q,Q\}\sim M$, then the
superspace identity (\ref{projection11}) must also hold. The supersymmetry
transformations (\ref{susytrans}) are unchanged, but we must have extra
constraints on the spinors in the theory.  

\bigskip

To conclude,  we reiterate the result. We have  found the general
supersymmetry transformations 
corresponding to any $N=1$ grading of the \po algebra. These gradings
all appear on nearly an equal footing. The
supersymmetry transformations are given by the expression
(\ref{susytrans}). In addition, if the
$\{Q,Q\}$ term generates $M_{\mu\nu}$ then the  expression (\ref{proj})
must hold in order to preserve the coset construction, although the
supersymmetry transformations are thus unaffected by this. For each form
$\Pi^{\mu_1\dots\mu_p}$ there are four possibilities, depending on
whether or not the superspace demands a projection of the spinors and
whether $\sigma_p=0$ or 1.

\section{The superspace identity}
\setcounter{equation}{0}
We now discuss the superspace identity (\ref{projection11}) for which
we must  define a restricted subclass of
spinors such that
\be\label{projectionorig}
\psi_\alpha(\Gamma^{\mu\nu})^{\alpha\beta}\phi_\beta=0\,\,\,\,\,\,\,\forall\mu,\nu\,.
\ee
These equations must be satisfied for all pairs $\phi,\psi$ and 
may be thought of as defining  a class of
spinors. This defining 
relationship is a completely covariant expression, and a theory involving such
spinors would therefore be Lorentz invariant. It is an interesting question to
ask which subsets of the space of spinors satisfy this equation. 
 We are  used to dealing
with Dirac or Majorana spinors, so we now investigate the relationship
between these and the new class of spinors.  Can a spinor satisfying
(\ref{projectionorig})  be obtained via a projection of a Dirac spinor 
\be\label{mm0}
\theta_D^\alpha\rightarrow \psi^\alpha={{\cal{P}}^\alpha}_\beta\theta_D^\beta\,,
\ee
where $\theta_D$ is a Dirac spinor? 
Since the equation (\ref{projectionorig}) must be satisfied for 
every 
$\phi$ and $\psi$  we discover  that the problem is equivalent to finding
projectors such that
\be\label{projection2}
{\widetilde{{\cal{P}}}}C{(\Gamma^{\mu\nu})}{{\cal{P}}}=0\,\,\,\,\,\,\,\forall\mu,\nu\,,
\ee
since  a matrix is orthogonal with respect to all vectors of
the appropriate dimension if and only if it  is the zero
matrix, even if the vectors are Grassmann-odd valued. 

To see that (\ref{projection2}) must hold we merely need choose a non-zero
spinor $\phi$ in the 
expression $ {{\cal{P}}^\alpha}_\beta\psi^\beta C_{\alpha\rho}{(\Gamma^{\mu\nu})^\rho}_\gamma{{\cal{P}}^\gamma}_\delta\phi^\delta$. We can then
always construct a  spinor $\psi$ for which this expression is
not zero. We must therefore impose the restriction
(\ref{projection2}).

We may ask what the rank of the projectors ${\cal{P}}$ must be. To
answer this question we note that the equation 
(\ref{projectionorig}) is in fact a {\it linear} constraint. Since the
identity must hold for all spinors, the spinors
$\lambda_1\psi+\lambda_2\phi$ must also be orthogonal to $\phi$ and
$\psi$ for any
real numbers $\lambda_1,\lambda_2$. We therefore must restrict the
spinors to lie in some vector subspace of the full Dirac spin
space. This implies that the number of spin degrees of freedom will be
some multiple of 2, hence the projectors ${\cal{P}}$ will be of rank
${\left(\frac{1}{2}\right)}^n$ for integer values of $n$. 
A similar set of identities to (\ref{projection11}) are used to define
the so called pure spinors \cite{pure_spinors}. The pure spinor
relationship, however, admits non-linear solutions, which leads to
unusual degrees of freedom, unlike the case here.

We now wish to investigate the  construction of  some of the projectors which
satisfy (\ref{projection2}), to discover for which spaces we may
define the new spinors.  There are two sub-cases
to consider: those for which ${\cal P}$ is 
invariant under $SO(S,T)$ rotations, which we shall call Lorentz
invariant,  and those for which it is not.

\subsection{$SO(S,T)$ invariant cases}

There is essentially only one projector which is invariant under
$SO(S,T)$ rotations: the Weyl projector
\be\label{proj2}
{\cal{P}}=\frac{1}{2}(1+\Gamma^{D+1})\,,
\ee
where  $\Gamma^{D+1}=\Gamma^{1}\dots\Gamma^{D}$. In order that
${\cal{P}}^2={\cal{P}}$ we must have 
$(\Gamma^{D+1})^2=+1$. We must additionally choose $D$ to be even, so
that $\Gamma^{D+1}$ is not proportional to the identity matrix, to
ensure that ${\cal{P}}$ is a non-trivial projection matrix.  
However, ${\cal{P}}$ does not always satisfy
the identity  (\ref{projection2}), and it
is not always possible, therefore, to construct an $SO(S,T)$ invariant
superspace if $\{Q,Q\}$ generates $M_{\mu\nu}$. In fact, we have the
following result

{\underline{Lemma}}

Given an irreducible Majorana representation of the $D$ dimensional
Clifford algebra, the identity (\ref{projection2}) is satisfied for
${\cal{P}}=\frac{1}{2}(1+\Gamma^{D+1})$ if and only if
$T$ is odd, 
$D\mod 4=2\,$ and $(S-T)\mod 8 =0$. 

{\underline{Proof}}

Since we have a Majorana representation of the Clifford algebra we
take the gamma matrices to be real. Suppose that $\tilde{\Gamma}^\mu=kC\Gamma^\mu C^{-1}$. We may always
choose a basis for the gamma matrices such that
${(\Gamma^{\mu_t})^{\dag}}=-\Gamma^{\mu_t}$ for $\mu_t$ a timelike index
and ${(\Gamma^{\mu_s})^{\dag}} =+\Gamma^{\mu_s}$ for $\mu_s$ spacelike \cite{nieuwenhuizen}. 
Then we have that 
\be
\Gamma^{\mu_t}C=-{k}C\Gamma^{\mu_t}, \hskip1cm
\Gamma^{\mu_s}C=+{k}C\Gamma^{\mu_s}\,,
\ee
Using these expressions we find that 
\be
C\Gamma^{1}\dots\Gamma^{D}=(-{k})^T({k})^S\Gamma^{1}\dots\Gamma^{D}C\,.
\ee
We also have
\ben
(\Gamma^1\dots\Gamma^D)^\sim&=&
\tilde{\Gamma}^D\dots{\tilde{\Gamma}}^1=(-1)^T\pi\Gamma^1\dots\Gamma^D\nonumber\\
\pi&=&(-1)^{\left[(D-1)+(D-2)+\dots 2+1\right]}\nonumber\\
&=& \left\{ \begin{array}{ll}
+1& \mbox{ if } D\mod 4=0,1\\-1 & \mbox{ if } D\mod 4=2,3
\end{array}\right. \\ 
\een

We now consider the matrix equation
\ben\label{matrixeqn}
\tilde{\cal{P}}\Gamma^{\mu\nu}C{\cal{P}}&=&\frac{1}{2}\left(1+(-1)^T\pi\Gamma^1\dots\Gamma^D\right)\left(1+(-{k})^T({k})^S\Gamma^1\dots\Gamma^D\right)\Gamma^{\mu\nu}C\nonumber\\
&=&\frac{1}{2}\left(1+((-{k})^T({k})^S+(-1)^T\pi)\Gamma^1\dots\Gamma^D+
\pi^2(-1)^T({k})^D\right)\Gamma^{\mu\nu}C\nonumber\\
&=&0.
\een 
If this equation holds, then so does (\ref{projection2}).
For (\ref{matrixeqn}) to be true, we must have that, assuming that
$\Gamma^1\dots\Gamma^D$  is not proportional to the identity matrix, 
\be\label{id1}
(-{k})^T({k})^S+(-1)^T\pi= \pi^2(-1)^T({k})^D+1=0\,.
\ee
Evaluating all the possibilities we find that these equations are
satisfied only if
\newcounter{xxx}
\begin{list}
{(\roman{xxx})}{\usecounter{xxx}}
\item
$T$ is even and $S$ is odd, $D\mod 4=1$ and $C=C_-$
\item
$T$ is odd and $S$ is even, $D\mod 4=3$ and $C=C_+$
\item
Both $T$ and $S$  odd, $D\mod 4=2$ and $C=C_+$ or $C=C_-$. 
\end{list}
We must check which of these possibilities are consistent with the
Majorana condition, by referring to (\ref{eta}) and (\ref{tab}).
 We find  that  item (i) is always inconsistent and items (ii) and
(iii) only hold if $(S-T)\mod 8$ equals 1 and 0 respectively. We thus have that
the equation (\ref{matrixeqn}) is satisfied  for the projector
${\cal{P}}$ if and only if $T$ is odd, $D\mod
4=2,3$ and $(S-T)\mod 8=$0 or 1. We now recall that in odd dimensions 
$\Gamma^1\dots\Gamma^D$ is proportional
to the identity matrix, in which case the  spinor identity is never
solved.   This requires 
us to choose 
$D$ to be even.  Listing all the possibilities provides the result   
$\Box$.

\subsection{Non-$SO(S,T)$ invariant cases}\label{sectionp}

We now search for some general types of projector which satisfy the
superspace identity in 
a non-covariant fashion. Of course, the underlying theory is still
Lorentz invariant: The choice of projector is analogous to a
gauge choice.  We need not, therefore,   worry that we lose {\it manifest}
Lorentz invariance.

\subsubsection{Product projectors}

To begin with, we consider acting on the spinors with a product of
projectors of the form ${\cal{P}}_{(s,t)}=\frac{1}{2}(1+\Gamma_{(s,t)})$ where
$\Gamma_{(s,t)}$ is the 
product of $s$ spacelike and $t$ timelike gamma matrices.  Since
${\cal{P}}$ is to be a projector, we require that
$\Gamma_{(s,t)}^2=1$ and that  $\Gamma_{(s,t)}$ is not 
proportional to the identity. This is possible iff  $(s-t)\mod 8=0$.

 For a product of $n$ such
projectors, the left hand side of the superspace identity becomes
\be\label{L}
L=(1+{\widetilde{\Gamma}}_{(s_1,t_1)})\dots
(1+{\widetilde{\Gamma}}_{(s_n,t_n)})C_{\pm}^{-1}\Gamma^{\mu\nu}(1+\Gamma_{(s_n,t_n)})\dots
(1+\Gamma_{(s_n,t_n)})\,,
\ee
which becomes
\ben
L_+&=&C_+^{-1}(1+(-1)^{t_1}\Gamma_{(s_1,t_1)})\dots
(1+(-1)^{t_n}\Gamma_{(s_n,t_n)})\Gamma^{\mu\nu}(1+\Gamma_{(s_n,t_n)})\dots
(1+\Gamma_{(s,t)})\nonumber\\
L_-&=&C_-^{-1}(1+(-1)^{s_1}\Gamma_{(s,t)})\dots (1+(-1)^{s_n})\Gamma_{(s_n,t_n)})\Gamma^{\mu\nu}(1+\Gamma_{(s_n,t_n)})\dots
(1+\Gamma_{(s,t)})\hskip1cm\,,
\een
for the two choices of the charge conjugation matrix. Evaluating all
the 
possibilities, we find that
$L=0$  if and only if
\begin{enumerate}
\item
The sets of
gamma matrices  labelled by $(s_i,t_i)$ for  $i=1,\dots ,n$ form a
partition of 
  $\{\Gamma^1,\dots,\Gamma^S,\Gamma^{S+1},\dots,\Gamma^{(S+T)}\}$.
\item
$(-1)^{t_1}\dots(-1)^{t_n}=-1$ for $C_+$ and
$(-1)^{s_1}\dots(-1)^{s_n}=-1$ for $C_-$. 
\item
All of the matrices $\Gamma_{(s_i,t_i)}$ for $i=1,\dots,n$ commute
with each other. 
\item
The matrices  $\Gamma_{(s_i,t_i)}$ are all independent. 
\end{enumerate}
Studying these constraints for $D\le 14$ and $T\le 3$ provides 
a single non-Weyl solution. This is given by the decomposition
(8,0)(1,1)(1,0) in signature (10,1). 
Although many other decompositions appear to fit this picture, they
are inconsistent for the final  reason on the previous list. 
 By substituting an
explicit representation of the (10,1) gamma matrices, it is easy to
show that the rank of the projector corresponding to (8,0)(1,1)(1,0)
is $\frac{1}{4}$.

\subsubsection{Tensor product projectors}
We now try to determine the rank of the solution of the superspace
identity in certain signatures, by considering tensor product
projectors of the form
\be
{\cal{P}}={\cal{P}}_1\left(I_2\otimes{\cal{P}}_2\right)\,,
\ee 
for a projector  ${\cal{P}}_1$ of dimension $2^{[D/2]}\times
2^{[D/2]}$ and a smaller  projector
${\cal{P}}_2$ of dimension $2^{[D/4]}\times
2^{[D/4]}$. Of course, since we wish to deal with
Majorana spinors, 
we must require that Majorana spinors exist on the subspace acted on
by ${\cal{P}}_2$ as well as in the full space.

We consider spacetimes with both spatial and temporal directions. We
may then write 
\be\label{rep2}
\Gamma_0=\pmatrix{0&1\cr -1&0}\,,
\hskip1cm\Gamma_D=\pmatrix{0&1\cr1&0}\,, \hskip1cm 
\Gamma_p=\pmatrix{\gamma_p&0\cr 0&-\gamma_p}\,p=1\dots D-1\,;
\ee
the charge conjugation matrices are given by
\be\label{rep2c}
C_+=\pmatrix{0&c_-\cr c_-&0}\,,\hskip1cm C_-=\pmatrix{0&c_+\cr
-c_+&0}\,,
\ee
with $c_\pm$ being the charge conjugation matrices for the
$\gamma_1\,\dots,\gamma_{D-1}$. 
We now reduce the problem to a similar one  in a lower dimension by
employing  
the projector 
\be
{\cal{P}}=\frac{1}{2}(1+\Gamma_0\Gamma_D)(I_2\otimes{\cal{P}}_2)\,,
\ee
where ${\cal{P}}_2$ is a matrix projector of the same dimension as the
$\gamma_p$, and $I_2$ is the two dimensional identity matrix. We then
find that the problem is equivalent to solving
\be\label{newid2}
{\widetilde{\cal{P}}}_2 c_\mp\gamma^p{\cal{P}}_2=0\,,
\mbox{ for } C=C_\pm \mbox{ respectively }\,.
\ee
If  ${\cal{P}}_2=\frac{1}{2}(1+\gamma_a\dots\gamma_{D-1})$ is a
projection operator, which is the case iff $(S-T)\mod 8=0$,  then  
the equation (\ref{newid2}) has a solution iff 
$T$ is an even number (and non-zero, of course). 
This gives the total rank of the projector ${\cal{P}}$ to be one quarter in the
cases $T\mod 8=S\mod 8$. 

We may now repeat the reduction procedure on the matrix identity
(\ref{newid2}) 
involving the lower dimensional gamma matrices $\gamma_p$, if we have
that $T,S\geq 2$. In an exactly 
analogous way, if  we consider the  projector
\be
{\cal{P}}=\frac{1}{2}(1+\Gamma_0\Gamma_D)I_2\otimes\left(\frac{1}{2}\left(1+\gamma_t\gamma_s\right)I_2\otimes{\cal{P}}_4\right)
\ee
we find that the surviving piece of the identity is 
\be\label{maria}
{\widetilde{\cal{P}}}_4 C {\cal{P}}_4=0\,,
\ee
where ${{\cal{P}}}_4$ is a projection matrix of dimension
$2^{[D/8]}\times 2^{[D/8]}$, and $C$ is the appropriate charge conjugation
matrix. The equation (\ref{maria}) always has a solution for a rank
$\frac{1}{2}$ 
projector ${\cal{P}}_4$. Thus the complete projector ${\cal{P}}$
required to satisfy 
 the full superspace identity is of rank $\frac{1}{8}$.

\bigskip
We have presented various methods for constructing projectors
which satisfy the superspace identity (\ref{proj}). There may, of
course, be other projectors which satisfy the identity
(\ref{projection2}), although we 
have considered  the obvious constructions of such objects. We present the
signatures with $S\geq T$ and $D \leq 14$ for which a solution has
been found for real 
spinors, as 
well as the rank of the appropriate projection matrix
\be
\begin{array}{|c|c|c|}\hline
D&(S,T)&\mbox{Rank}\\\hline
2&(1,1)&1/2\\
6&(3,3)&1/2\\
10&(9,1),(5,5)&1/2\\
11&(10,1)&1/4\\
12&(10,2),(6,6)&1/4\\\hline
\end{array}
\ee
In addition, we also have that there exist projectors of rank
$\frac{1}{8}$ in the cases that a space of signature $(S,T)$ admits
Majorana spinors, and that $S,T\geq 2$. Some examples of interest in
twelve, thirteen and fourteen dimensions are
$(S,T)=(9,3), (10,3), (11,2), (11,3)$.

\section{$p$-brane  actions}
\setcounter{equation}{0}

We now turn to the question of $p$-branes moving in  general
$\mbox{sio}_k$ 
 superspaces. From studies of the usual $\mbox{sio}_1$ \po
 supersymmetry the existence of different types of $p$-brane has
 emerged. The hope is that certain branes will have a more natural
 description in the new types of superspace and that we will discover
more branes by employing this more comprehensive description of supersymmetry. 
 A $p$-brane in $D$ dimensions is a
$p+1$-dimensional brane manifold embedded  in a $D$ dimensional
 spacetime manifold. The $p$-branes we shall discuss correspond to fundamental
 objects, such as Green-Schwarz superstrings,  as opposed to
 solitonic $p$-branes. Usually the
spacetime is taken to be of signature $(D-1,1)$, and the brane
to be of signature $(p,1)$ but we shall lift
these restrictions. The `spacetime' and the $p$-brane will be of
signature $(S,T)$ and $(s,t)$ respectively, so that $D=S+T$ and
$p+1=s+t$ to fit in with the usual definition of a $p$-brane for
$t=1$; to be more precise one could call the brane an 
$(s+t)$-brane.  In order that the construction
be classically stable, which corresponds to the absence of ghosts
quantum mechanically, we require that $t=T$ \cite{duff_blencowe}.  This removes the
possibility of negative norm states propagating in directions
transverse to the brane.

We now assume the principle of least action for a $p$-brane moving in a
 general \po superspace to produce a set of  $p$-brane
actions. Recall that the bosonic part of the superspace can be
 parametrised by the terms 
$\{X^{\mu_1\dots\mu_n}\}$ where $n\in P\subseteq\{1,3,4,\dots,D\}$. We thus
may write down the actions
\be\label{action}
S_P=\int \,d^{p+1}\xi[\det(\sum_{n\in P}\Pi_i^{\mu_1\dots\mu_n}\Pi_j^{\nu_1\dots\nu_n}\eta_{\mu_1\nu_1}\dots\eta_{\mu_n\nu_n})]^{\frac{1}{2}}\,,
\ee
where $\Pi_i^{\mu_1\dots\mu_n}$ are the pullback of the forms
(\ref{oneforms}) to the worldvolume of the brane, which has
coordinates $\xi_1,\dots,\xi_{p+1}$.  
These  actions are manifestly spacetime supersymmetric  under the
action   
of the supergroups corresponding to $\mbox{sio}_{\tilde{k}}$, since
 they are constructed from invariant forms, and are
generalisations  of the standard Dirac type  actions to include higher spin
 gauge fields on the worldsheet. 

It is  a
matter of interpretation to decide which actions are physically
relevant to {\it{structureless}}\,(fundamental) $p$-brane
propagation. Under the superspace rescaling 
$X^{\mu_1\dots\mu_n}\rightarrow \Omega^{n}X^{\mu_1\dots\mu_n}$, 
each term in the sum 
\be\label{sum}
(\sum_{n\in P}\Pi_i^{\mu_1\dots\mu_n}\Pi_j^{\nu_1\dots\nu_n}\eta_{\mu_1\nu_1}\dots\eta_{\mu_n\nu_n})\,,
\ee
scales differently. We therefore suppose that the fundamental actions
are written as
\be\label{fundaction}
S_n=\int
 \,d^{p+1}\xi[\det(\Pi_i^{\mu_1\dots\mu_n}\Pi_j^{\nu_1\dots\nu_n}\eta_{\mu_1\nu_1}\dots\eta_{\mu_n\nu_n})]^{\frac{1}{2}}
\equiv\int
 \,d^{p+1}\xi[\det(\Pi_i^{(n)}.\Pi_j^{(n)})]^{\frac{1}{2}}\,.
\ee
These actions can be rewritten in the Howe and Tucker form
 \cite{howe_tucker} as follows
\be{\label{ht}}
S_n=\int\,d^{p+1}\xi\left(\sqrt{|g|}\,g^{ij}\Pi_i^{(n)}.\Pi_j^{(n)}-\frac{1}{2}(p-1)\right)\,,
\ee
where $g_{ij}$ is the induced metric on the worldsheet.
Thus for $n>0$ we have  a natural way in which to define the
action for one of 
the $n$-form charges propagating on the worldsheet. This idea may be
of use if we wish to add degrees of freedom to the $p$-brane problem
by adding higher spin fields on the brane. To see that this  is
consistent, we note that the equation of motion of the $p$-form is
simply
\be
dF=0\,,\hskip0.5cm F=d\Pi^{(p)}\,,
\ee
as one would expect. This is a pleasing result, since it was
obtained via the principle of least action for a manifold invariant
under the action of a generalised \po supergroup.


\subsection{Wess-Zumino terms}
\subsubsection{$S_1$ case}
The standard $p$-brane actions may be augmented with an additional
piece, called the Wess-Zumino action. This  term is added to the basic
action $S_1$ because it contains an antisymmetric tensor field which
provides a
coupling of the brane to the local supergravity theory. 
We  define the Wess-Zumino
term to be 
of the form  
\be\label{wz}
S_{WZ}=-\int\,d^{p+1}\xi\left(\frac{1}{(p+1)!}\epsilon^{i_1\dots
i_{p+1}}B_{i_1\dots i_{p+1}}\right)\,,
\ee
where $B_{i_1\dots i_{p+1}}$ are the components of the pullback to the
brane  of  a super $(p+1)$-form, 
$B$,  which is the potential for a super $(p+2)$-form  $H=dB$. For the
action $S_1$, parametrised by $(X^\mu,\theta)$,  consistency requires  the
Wess-Zumino 
integral to scale in the same way as $S_1$  under the
superspace rescaling 
$X^\mu\rightarrow\Omega X^\mu,
\theta^\alpha\rightarrow\Omega^\frac{1}{2}\theta^\alpha$.
For a given $p$ we can then define $H$ to be
\be\label{db}
H=\Pi^{\mu_1}\dots\Pi^{\mu_{p}}d\bar{\theta}\Gamma_{\mu_1\dots\mu_p}d\theta\,,
\ee
where the forms $\Pi^\mu$ are defined in (\ref{oneforms}). For this
not to vanish identically, $(\Gamma_{\mu_1\dots\mu_p})_{\alpha\beta}$
must be symmetric in the spinor indices, since the $d\theta$ are
commuting variables as the $\theta$ are Grassmann odd. 
Taking the
exterior derivative of the forms $\Pi^\mu$  we find that
\be
d\Pi^\mu=\sigma_1
d\theta_\alpha(\Gamma^\mu)^{\alpha\beta}d\theta_\beta\,. 
\ee
Since $\sigma_1=0$ automatically if $\Gamma^\mu$ is antisymmetric in
the spinor indices, $d\Pi^\mu=0$ iff  $\sigma_1=0$. 
   As 
$H=dB$ we must check that  $dH=0$ for consistency. If $\sigma_1=1$ then
for $H$ to be closed it is well known that  
then  we must have 
that \cite{branescan} 
\be
D-p-1=\frac{nN}{4}\,,
\ee
where $n$ is the number of spin degrees of freedom of the spinors and
$N$ is the 
number of supersymmetry generators. If $\sigma_1=0$ then $H$ is 
automatically closed, due to the triviality of the superspace forms. 
This differing character of the superspace forms for the two different
types of supersymmetry also 
gives rise to two different types of $B$. Up to total derivatives,
for the standard $\sigma_1=1$ case, $B$ may be shown to be
\cite{townsend:trieste} 
\be\label{b1}
B=\frac{(-1)^p}{2(p+1)!}(d\bar{\theta}\Gamma_{\mu_1\dots\mu_p}\theta)\left[\sum_{r=0}^p(-1)^r\left({p+1}\atop{r+1}\right)\Pi^\mu_p\dots\Pi^\mu_r+1(d\bar{\theta}\Gamma^{\mu_r}\theta)\dots(d\bar{\theta}\Gamma^{\mu_1}\theta)\right]\,.
\ee
For the $\sigma_1=0$ superspaces the situation is much simpler,
since all the forms  in the problem are exact. We find that 
\be\label{b2}
B=\frac{(-1)^p}{(p+1)!}dX^{\mu_1}\dots
dX^{\mu_p}(d\bar{\theta}\Gamma_{\mu_1\dots\mu_p}\theta)\,,
\ee
which may be obtained from (\ref{b1}) by setting
$d\bar{\theta}\Gamma^\mu\theta=0$.  
It is well known that the action $S_1+S_{WZ}$ is invariant under a
local 
$\kappa$-symmetry transformation for the standard $\mbox{sio}_1$
superspace. We need to investigate whether or not this is the case for
all the superspaces for the action $S_1$. 
\subsubsection{$S_n$ case, for $n>1$}
One may always define the Wess-Zumino term for the action $S_1$, but
this is not always the case with  actions for higher values of $n$. If
one considers a theory with simply the coordinates
$\{X^{\mu_1\dots\mu_n},\theta\}$ then the Wess Zumino term may only be
constructed by the scaling argument if $p$ is some integer multiple of
$n$. We would then find that
\be
H=\Pi^{\mu_1\dots\mu_n}\dots\Pi^{\mu_{(kn-n+1)}\dots\nu_{kn}}d\bar{\theta}\Gamma_{\mu_1\dots\mu_{kn}}d\theta\,.
\ee
If we have a theory parametrised by a larger set of coordinates, then
it may  possible to construct more general Wess-Zumino
terms, which are not necessarily unique. 
If, on the other hand, $p$ is not an integer power of $n$ then we
cannot construct a 
Wess-Zumino term  without introducing fractional  powers of the
integrand to get the
correct scaling behaviour. 

We shall henceforth mainly
consider the traditional type $S_1$ action, since it incorporates 
the basic coordinates of spacetime,  although we shall look at
those which are invariant under the general types of \po
supersymmetry.

\section{Generalised supersymmetry $p$-branes}
\setcounter{equation}{0}

\subsection{$\kappa$-symmetry for  the $n=1$ $\mbox{sio}_1$,
$\mbox{sio}_2$ and $\mbox{sio}_{1,2}$
invariant actions} 
The $\kappa$-symmetry is a hidden symmetry which arises in the standard
formulation of $p$-brane actions. This symmetry, which  is an
interrelationship between  $S_1$ and the Wess-Zumino term  in the
full $p$-brane action, is a special local spinor transformation. In
order to  discuss $\kappa$-symmetry in  general, it is instructive
to first detail the transformation in the standard context. 
For $\mbox{sio}_1$ superspace, the $\kappa$ symmetry is defined to be
a local version of the supersymmetry transformations
\be\label{normalkappa}
\delta X^\mu=\bar{\theta}\Gamma^\mu\delta\theta\,,
\ee
for some local spinor $\delta\theta$.  If this is so then the
variation in the forms 
$\Pi^\mu$ are 
given by
\be
\delta\Pi_i^\mu=-2\delta\bar{\theta}^\alpha\Gamma_{\alpha\beta}^\mu\diff_i\theta^\beta\,,\hskip1cm
\delta\Pi_i^\alpha=\diff_i\delta\theta^\alpha\,.
\ee
In order to obtain invariance of the full action $S_1+S_{WZ}$ under
these changes, we must relate the Wess-Zumino  term, which has no
metric dependence, 
to the integral $S_1$. To do this we work with the
pullback to the brane of the gamma matrices
\be\label{gammarelns}
\Gamma_i=\Pi_i^\mu\Gamma_\mu\,,\hskip0.5cm
\{\Gamma_i,\Gamma_j\}=2g_{ij}\,.
\ee
We then obtain the relationship
\be
\epsilon^{i\dots jk}\Gamma_{\i\dots
j}=2\sqrt{-g}g^{kl}\Gamma_l\Gamma\,,
\ee
with 
\be
\Gamma=\frac{(-1)^{(p+1)(p+2)/4}}{(p+1)!\sqrt{|g|}}\Gamma_{i_1\dots
i_{p+1}}\epsilon^{i_1\dots i_{p+1}}\,.
\ee
The matrix $\Gamma$ has the properties that $\mbox{Tr}(\Gamma)=0\,,$
iff $p+1\neq D$, and that $\Gamma^2$ is the identity matrix. 
The full action is then seen to be invariant under the $\kappa$-transformation
\cite{duff_blencowe} if we make the choice 
\be
\delta\theta=(1+\Gamma)\kappa\,.
\ee
We may thus gauge away half the spinor degrees of freedom.  It has not
been proven that this is the only 
 hidden symmetry of these $p$-branes
actions, but it is  difficult to think of any other
possibility.

The question of a $\kappa$-symmetry for a superspace without torsion 
is not so obvious. At the most basic level, in constructing a local
fermionic symmetry  of a $p$-brane action,  we require
that the fermionic variation is cancelled by the bosonic variation,
with an additional constraint on the spinors used in the variation. 
The Wess-Zumino term for an  invariant action with no torsion is the
integral over the  brane of the $(p+1)$-form $B$, given by
(\ref{b2}). Since the superspace forms are trivial, there is no link
between the bosonic and fermionic coordinates, and the interplay which
occurs in the usual definition of $\kappa$-symmetry, 
(\ref{normalkappa}),  does not occur.  
Writing down the variation of the action for the $n=1$ case, and using
the expression 
(\ref{gammarelns}), we find that
\be
\delta S=\int d^{p+1}\xi\,\sqrt{g}g^{ij}\left(2\diff_i(\delta
X^\mu)\diff_jX^\nu+\delta\bar\theta\Gamma\Gamma_j\diff_i\theta\right)-\int\,
d\bar\theta\Gamma_{\mu_1\mu_2\dots\mu_p}d\theta(\delta X^{\mu_1}
dX^{\mu_2}\dots dX^{\mu_p})\,.
\ee
In order for this variation to vanish, we must have that $\delta
X^\mu=0$, in which case the constraint on the variation for the
$\theta$ terms becomes 
\be\label{kid}
\delta\bar\theta\Gamma\Gamma_i\diff_j\theta=0\,.
\ee
If $\theta$ is a general Dirac or Majorana spinor, then this equation
is satisfied iff $\delta\theta=0$, in which case we find that there is
no $\kappa$-symmetry. For the $\mbox{sio}_2$  superspace, however,  we
must work with a spinor projection, in which case 
(\ref{kid}) is satisfied for non-zero $\delta\theta$ iff
\be\label{kid2}
\widetilde{{\cal{P}}}C\Gamma\Gamma_i{{\cal{P}}}=0\,.
\ee
For such $\mbox{sio}_2$ superspaces we already have that
\be
\widetilde{{\cal{P}}}^\pm C\Gamma^{\mu\nu} {{\cal{P}}}^\pm=0\,,
\ee
for some orthogonal projectors ${{\cal{P}}}^\pm$.  This implies that
\be
\widetilde{{\cal{P}}}^\pm C\Gamma^{\mu\nu}
{{\cal{P}}}^\pm=C\Gamma^{\mu\nu}{{\cal{P}}}^\mp{{\cal{P}}}^\pm\,, 
\ee
Since  $\Gamma\Gamma_i$ is a sum of independent products of $p+2$
gamma matrices, then by considering the expressions
\be
{\widetilde{\cal{P}}}^\pm
C\Gamma^\mu\Gamma^\nu\Gamma^{\rho_1}\dots\Gamma^{\rho_{p}}
{\cal{P}}^\pm=C\Gamma^\mu\Gamma^\nu{\cal{P}}^\mp\Gamma^{\rho_1}\dots\Gamma^{\rho_{p}}{\cal{P}}^\pm\,,
\ee
we find that
\begin{enumerate}
\item
If ${\cal{P}}$ is Weyl then (\ref{kid2}) is satisfied iff $p$ is even
or zero.
\item
If ${\cal{P}}$ is not the Weyl projector then (\ref{kid2}) is only
satisfied if we take $p=0$. 
\end{enumerate}
It is therefore often the case that there is no $\kappa$-symmetry for the  
superspaces with $\sigma_1=0$. This should be no surprise.  For a
supersymmetry of the 
form $\{Q,Q\}\sim P$ there exist local supergravity theories for
dimensions eleven or less. The $\kappa$-symmetry is a local version of
the rigid supersymmetry transformations. For a supersymmetry of the form
$\{Q,Q\}\sim M$ there is no local {\it supergravity}
theory since the $P$ terms generate the diffeomorphisms which enable
us to couple supersymmetry to general relativity. Such a coupling does
not occur for $\mbox{sio}_2$ theories,  hence we should expect the
related 
$\kappa$-symmetry problem to be more subtle.

\subsection{Worldvolume supersymmetry}

In the previous section we presented superspacetime symmetric
actions. We shall now  construct Green-Schwarz $p$-brane actions for
all possible  values of 
$p$ and $(S,T)$. To do this we 
 enforce supersymmetry on the 
brane, in addition to spacetime supersymmetry. We  consider
supersymmetric theories generated by $\{P,M,Q\}$ for which  the brane fields
form  scalar 
supermultiplets, given by $(\theta^\alpha,X^\mu)$.  For
this type of 
worldvolume supersymmetry to occur we  require that the brane bosonic
and fermionic physical degrees 
of freedom match up. 
We have many possibilities for the degrees of freedom counting,
depending on whether there exists a $\kappa$-symmetry for the brane
action and whether or not it is necessary to take a projection of the
spinors. 
In addition to these considerations, we are interested in
physical spinor  degrees of freedom.  We 
must therefore go `on-shell'. The spinor equations of motion are really second
class constraints, and thus half the number of degrees of
freedom. The final spinorial degrees of freedom must match the $D-p-1$
transverse bosonic degrees of freedom, to produce the degrees of
freedom matching formula  
\be\label{matching}
D-p-1=\kappa R\frac{n}{2}N\,.
\ee
In this expression,  $\frac{n}{2}$ corresponds to the  physical
degrees of freedom of an 
on-shell Majorana spinor, $n$ being the real dimension of the Majorana
spinor. $N$ is the number of supersymmetry generators, which we shall
take to be 1. 
The value of $R$ is the rank of any projection made on the spinors,
and $\kappa$ is equal to one half or unity, depending on whether there
is or is not a $\kappa$-symmetry of the $p$-brane action. 
We shall investigate branes for which a Wess-Zumino action exists, to
provide a coupling of the brane to local theories. For
a given $p$-brane this requires $\Gamma^{\mu_1\dots\mu_p}$ to be
symmetric in the spinor indices. 
The parity of such matrices is determined by the equation \cite{hp:ftheory}
\be\label{parity}
\pi=\epsilon\eta^T(-1)^{\frac{T(T+1)}{2}}\left((-1)^T\eta\right)^i(-1)^{\frac{i(i-1)}{2}}\,,
\ee
where the choices of $\epsilon$ and $\eta$ are given in
(\ref{eta}), from which we find that
$\pi(\Gamma^{\mu_1\dots\mu_i})$ is given by
\be\label{paritytable}
\begin{array}{c|c|c}
&i=1&i=2\\ \hline
T=0&\epsilon\eta&-\epsilon\\ \hline
T=1&\epsilon&\epsilon\eta\\
\end{array}
\ee
This table is anti-periodic modulo 2 for both $i$ and $T$. For a
Majorana representation we must have that $\epsilon=1$ and $(S-T)\mod
8=0,1,2,6,7$.  For such cases
we give the 
sets
$k_\pm=\{i:(\Gamma^{\mu_1\dots\mu_i}C_\pm)^{\alpha\beta}=(\Gamma^{\mu_1\dots\mu_i}C_\pm)^{\beta\alpha}\}$
for  
each choice of the charge conjugation matrix
\be\label{egg}
\begin{array}{c|c|c}
T\mod 4&k_+&k_-\\ \hline
0&\{1,4,5,8,9,12,\dots\}&\{3,4,7,8,11,12,\dots\}\\ \hline
1&\{1,4,5,8,9,12,\dots\}&\{1,2,5,6,9,10,\dots\}\\ \hline
2&\{2,3,6,7,10,11,\dots\}&\{1,2,5,6,9,10,13,\dots\} \\ \hline
3&\{2,3,6,7,10,11,\dots\}&\{3,4,7,8,11,12,\dots\}\\
\end{array}
\ee

\subsection{Brane scans}

We shall say that a $p$-brane action exists if, for a given
anticommutator 
(\ref{newqq}), there exists $p$, $D$ and signature such that (\ref{matching})
holds. We also require that $\Gamma^{(p)}$ be symmetric in its spinor
indices so that we may define a Wess-Zumino action. Finally, all of
these requirement must be consistent with the definition of a charge
conjugation matrix for real spinors in the given signature.   In the
cases for which a projection of the spinors is made we 
present all of  the branes which are allowed from a Bose-Fermi matching
point of view. The explicit construction of possible projectors is found
 previously.

\subsubsection{$(\kappa,R)=({\scriptstyle\frac{1}{2}},1)$}
This is the usual case for a $p$-brane with $\mbox{sio}_1$
supersymmetry, producing the usual brane scan. It is noteworthy that
the scan is {\it not} invariant under the interchange of
$S$ and $T$, given a fixed metric convention. The reason for this is
that the symmetry properties of products of gamma matrices vary
according to whether $T\mod 4=0,1,2,3$,  as may be seen in
(\ref{egg}). For example, we may define a 2-brane in signature
$(S,T)=(3,1)$, but no 2-brane exists in signature $(S,T)=(1,3)$. This
type of behaviour is present in all the brane scans. For some  
Minkowski dimensions of importance, namely (1,1), (5,1), (9,1), this
problem does not arise. Conversely,  we see that supersymmetric
physics  in (10,1) is 
not strictly  equivalent to that  in (1,10),  since $10\mod 4\neq
1\mod 4$ implies a different superalgebra structure.

\subsubsection{$(\kappa,R)=({\scriptstyle\frac{1}{2}},{\scriptstyle\frac{1}{2}})$}
Majorana-Weyl branes in $\mbox{sio}_1$ superspace, such as the $N=2$ type
IIA and type IIB superstrings  fall into this category. 
These values of $\kappa$ and $R$ would  include  Majorana-Weyl branes
propagating in the 
\siot superspace for which a $\kappa$-symmetry exists:  however, this  requires
$p$ to be even or zero, and   there are no such
solutions.

\subsubsection{{$(\kappa,R)=(1,{\scriptstyle\frac{1}{4}})$}}
These are branes which propagate in the  $\mbox{sio}_2$ superspace. The
solutions with no $\kappa$-symmetry are 
\be\label{table:kc}
\begin{array}{|c|c|c|c|c|c|c|}\hline
S=10&&{6^-}&{3^+}&&\\
9&&5^-&{6^\pm}&{3^+}&\\
8&&&{5^-}&&\\
7&&&&&\\\hline
T&0&1&2&3&4\\ \hline
\end{array}
\ee 
This table repeats for $(S,T)\rightarrow (S-4,T+4)$. Projectors for the
signatures $(S,T)=(9,1), (10,1), (10,2), (6,6)$ 
and $(1,10)$  have been explicitly constructed in the previous
sections. The other values are the only others  which are allowed by
symmetry and 
degrees of freedom matching. These would certainly occur in an $N=2$
brane scan with spinors projected by a rank $\frac{1}{8}$
projector. The three-brane in $(10,2)$ corresponds 
to the super 
$(2+2)$-brane \cite{hp:ftheory}, which reproduces the type
IIB string and the M-theory 2-brane after dimensional reduction. The
(10,1) six-brane is a new type of brane in eleven dimensions. Although
\sioo supersymmetry does not permit such an object, it naturally
occurs within the context of the generalised supersymmetry. This
brane could then provide a higher dimensional origin to the type IIA
six-brane.

\subsubsection{{$(\kappa,R)=(1,{\scriptstyle\frac{1}{2}})$}}
These are \siot branes with no $\kappa$-symmetry defined in an
explicitly Lorentz invariant way using 
Majorana-Weyl spinors. This requires us to choose $p$ to be odd. The
solutions are given in the following table, which is symmetric in $(S,T)$
\be\label{table:c}
\begin{array}{|c|c|c|c|c|}\hline
S=9&\,\,&1^\pm&&\\
3&&&&3^\pm\\\hline
T&0&1&2&3\\\hline
\end{array}
\ee
The 1-brane we obtain in this picture is very interesting,
as it is the simple supersymmetry analogue of the Green-Schwarz string
in \sioo superspace. The corresponding string action is constructed
from flat one-forms, and is thus in some sense a trivially supersymmetric
object. Of course, there is no reason why the \siot string  should be
the same object as the 
traditional string, and merely points to the existence of a  new type
of 1-brane. 

\subsubsection{{$(\kappa,R)=(\scriptstyle{\frac{1}{2}},{\scriptstyle\frac{1}{4}})$}}
There are no solutions for these degrees of freedom

\subsubsection{{$R={\scriptstyle\frac{1}{8}}$}}
These degrees of freedom only  give us a $7^+$ brane in signature
(9,3), if we have no
$\kappa$-symmetry. Otherwise, we find that there are no consistent
branes.

\subsection{Discussion}
\setcounter{equation}{0}

We have  presented all the possible branes with worldvolume scalar
supermultiplets 
propagating   in general \po superspaces. 
It should be possible to extend the analysis to consider $p$-branes
with higher spin fields on the worldsheet, which would render the
generalised supersymmetry arguments applicable to $D$-branes,  although we do not consider
this problem here. Given the existence of a particular $p$-brane in
the previous tables it is
 a simple 
matter to construct associated action, using the results from the
previous sections.  All the underlying theories to
these branes are fully covariant, although some of the actions may
need a non-covariant projector for their explicit description. 
We present the 
new branes for which we can construct {\it explicit} superspace projections
for low $T$. They all occur for superspaces with $\{Q,Q\}=M_{\mu\nu}\Gamma^{\mu\nu}$
\be
\begin{array}{|c|c|c|}\hline
(S,T)&p&(\kappa,R)\\\hline
(10,2)&3^+&(1,{\scriptstyle\frac{1}{4}})\\\hline
(9,3)&7^+&(1,{\scriptstyle\frac{1}{8}})\\\hline
(10,1)&6^-&(1,{\scriptstyle\frac{1}{4}})\\\hline
(9,1)&1^\pm&(1,{\scriptstyle\frac{1}{2}})\\
(3,3)&3^\pm&\\\hline
\end{array}
\ee

Several interesting points are raised by these results, not least that
it is possible to formulate brane theories using a simple
supersymmetry, which arises from the most natural fermionic extension
of the \po algebra.  In such a scenario we
find  a description of a superstring in a flat superspace with no torsion.  We
also find that 
generalised supersymmetry provides only a small number of additional
fundamental branes which must propagate in a maximal spacetime of
dimension twelve. Of course this maximal dimension arises if one
considers scalar supersymmetry on the brane, and may be altered in
more general scenarios. Twelve dimensional theories have been employed
recently to answer some problems associated with lower dimensional
physics \cite{twelve_dimensions,ftheory}. 
Dimensional reduction of the \siot threebrane and  six-brane 
should in principle be able  reproduce many of the fundamental branes in lower
dimensional string 
theory and M-theory, the necessary superspace torsion
being introduced upon compactification as in \cite{hp:ftheory}. 
It is interesting to note that in dimensions ten and eleven,
Minkowskian signature branes are singled out for all the types of
supersymmetry, whereas in the maximal twelfth dimension we see a
Kleinian space with two timelike directions appearing. This ties in
well with much of the other work on twelve dimensions for which a
signature of (10,2) is necessary. In order to dimensionally reduce the Kleinian
branes down to Minkowski signature branes we would  need to reduce on a
Lorentzian torus, as in the case of the twelve dimensional
F-theory. It has been suggested 
that the F-theory construction 
merely makes use of auxiliary extra dimensions and that 
only 10 of them are `real'. Generalised
supersymmetry, however, points to intrinsically twelve dimensional
theories, independent of any lower dimensional considerations.

\section{Conclusion}
\setcounter{equation}{0}

We have considered the most general supersymmetric extensions of the
\po algebra of spacetime and constructed the associated
superspaces. This analysis reveals that there are in fact two distinct
classes of supersymmetry. It also shows that we must in general  work 
with a new class of spinors, of which Weyl spinors are a special
case. 
Constructing $p$-branes in these new superspace backgrounds
produces additional  points on the brane scan which are not present for the
usual description of supersymmetry. The methods used are natural and
seem to provide some new ideas in string theory. The new class of supersymmetry
does not provide many new points on the brane scan, but 
produces a different set of $p$-branes to the ones usually
obtained. These $p$-branes create, in a sense, a  complete 
brane scan in 
dimensions ten and eleven, and give a higher dimensional origin for
some of the branes in such dimensions.  This could have corollaries
for  the
supergravity theories in which the $p$-branes are solutions.  It
should  therefore be of interest to investigate the implications of
generalised supersymmetry in other aspects of string theory. In
particular, it would seem to be  an interesting problem to investigate the
local version of the $\mbox{sio}_2$ supersymmetry.  This type
of theory would 
be a twelve dimensional `supergravity' theory. Possible forms of such
a theory have been 
discussed previously \cite{twelve_dimensional_supergravity}. Obviously
these could  not be of
the same form as the true supergravity theories in lower dimensions
since the anticommutator of the spinor generator with itself does not
produce a translation. The existence of such a local theory in twelve
dimensions would doubtless be related in some sense to supergravities
in lower dimensions. The discovery of such a theory could very well,
therefore, aid our understanding of the Minkowski space signature
theories we are ultimately interested in. 

Although some of the corollaries of generalised supersymmetry
presented in this paper may seem to be at odds 
with standard 
$p$-brane   
folklore, such as the existence of an eleven dimensional six-brane and
the twelve dimensional three-branes, they all stem from the
assumption that supersymmetry is a 
fermionic extension of the \po algebra, and that all such extensions
should basically be equivalent. Reassuringly, however,  these new
branes do seem to 
provide a unifying scheme for many of the previously known $p$-brane
theories.  We have become accustomed to the idea
that no one particular brane should be singled out as fundamental in
theories of extended objects; perhaps the same idea should now be
applied to supersymmetry.

\section{Acknowledgements}
I would like to thank Neil Lambert, Mike Green and Malcolm Perry for
useful discussions and comments. This work was supported by the
EPSRC.

\end{document}